# Thermal noise driven computing

Laszlo B. Kish[a]

Texas A&M University, Department of Electrical and Computer Engineering, College Station, TX 77843-3128, USA; email: Laszlo.Kish@ece.tamu.edu

(July 2, 29; 2006)

**Abstract**. A new type of computing, where thermal noise is the information carrier and the clock in a computer, is proposed. The energy requirement/dissipation are studied in a simple digital system with *zero threshold voltage*, when the thermal noise is equal to or greater than the digital signal. In a simple realization of a thermal noise driven gate, the lower limit of energy needed to generate the digital signal is $\approx 1.1*kT/bit$. The arrangement has potentially improved energy efficiency and it is free of leakage current, crosstalk and ground plane electromagnetic interference problems. Disadvantage is the larger number of required elements.

---

[a] Until 1999: L.B. Kiss

Recently, it has been shown [1-4] that modulated thermal noise (Johnson noise) can be a special type of information carrier and it can be used for stealth communication [1], totally secure non-quantum communication [2,3], and the realization of totally secure classical networks [4]. These results inspire a further question:

*If Johnson noise is such a peculiar information carrier, can it perhaps be applied to data processing and computing, too?*

Though, we do not know the full answer to this question, in this Letter, we shall show a potential application of Johnson noise to reduce the problems of energy dissipation, leakage current, crosstalk and ground plane electromagnetic interference (EMI) in microprocessors. This idea may first look strange because the ultimate lower limits of energy dissipation in computers is dictated by Johnson noise, see [5-8] and references therein. However, here we attempt to put thermal noise to work for us by driving the logic devices. We shall analyze how much is the information channel capacity of a particular realization of the thermal noise driven digital information channel and what is the lower limit of the energy requirement of transferring information through such a channel. Our other inspiration is the fact that neural signals are stochastic, which indicates that the brain is using *noise as an information carrier* [9], and at the same time the brain is an extremely energy efficient signal processor [10]. It is tempting to assume that the great energy efficiency of the brain is somehow related to the stochastic nature of neural signals [10].

Note that John von Neumann [11]; Forshaw and coworkers [12,13]; and Palem and coworkers [14,15] have been proposing efficient ways of working with probabilistic switches, which are noisy digital logic units with relatively high error probability. Palem and coworkers have pointed out that this may be a way to reduce power dissipation [14,15]. However, though these approaches can be relevant to future developments of the ideas outlined in the present paper, they are very different from our present approach. The system we propose is working in the regime of huge error probability, $\approx 0.5$, with *zero logic threshold voltage* and in the *sub-noise* signal amplitude limit, out of the range used by others. In the thermal noise driven computer the voltage in the channel is basically a noise and the *statistical properties of this noise carry the information*. Moreover, we base our study on *Shannon's channel coding theorem* because we believe that this theory is the proper application tool to characterize the ultimate information channel capacity and the energy efficiency in the relevant digital channels.

Therefore, the *thermal noise driven computer* is a computer where (most of) the logic gates are driven by a very small DC digital signal voltage $U_s$, which is zero in the "low" logic state, $U_{sL} = 0$, and it is equal to or less than the effective Johnson noise voltage $\sigma$ in the "*high*" logic state $U_{sH} \leq \sigma$. This situation will produce an extremely high error rate, close to the limit of 0.5 error probability in the digital channel, which is the case of zero information content. Thus the information channel has very low information content however, as we shall show, the energy requirement of creating the digital signal is also very low. Concerning the energy efficiency of such a computer, the important question is the energy requirement of handling a single bit (*Joule/bit*). We shall calculate the *lower limit* of this energy requirement.

The highest possible information rate in a digital channel is given by Shannon's channel coding theorem:

$$C_{dig} = f_c \left[ 1 + p \log_2 p + (1-p) \log_2 (1-p) \right] , \qquad (1)$$

where $C_{dig}$ is the information channel capacity, $f_c$ is the clock frequency, $p$ is the probability of correct bit and $1-p$ is the error probability, see Figure 1. Though Eq. (1) provides the exact value of the information channel capacity, it does not show us what kind of encoding is needed to approach this limit.

The lowest (second) order of the Taylor expansion of Eq. 1 around the value $p = 0.5$ yields

$$C_{dig}\big|_{p\approx 0.5} = (\Delta p)^2 \frac{2}{\ln 2} f_c \ , \tag{2}$$

where $\Delta p = p - 0.5$. The parabolic approximation given by Eq. (2) is very accurate in the interesting range ($p \approx 0.5$). Even at $p = 1$, which is far out of our range and where the inaccuracy of approximation is maximal; the relative inaccuracy is less than 30%. We will see that not only $C_{dig}$ but also the electrical power requirement to generate the signal is scaling with $(\Delta p)^2$ in the interesting range. Therefore, when the error rate is approaching 0.5 and the information content goes to zero, the energy requirement of generating a bit converges to a fixed value, where this value may depend on the realization of the computer.

For a possible realization of the thermal noise driven computing, let us suppose that a weak digital signal $U_s(t)$, where the "*low*" and "*high*" levels satisfy $U_{sL} = 0$ and $0 < U_{sH} \leq \sigma$, drives a simple *comparator* [16] with zero reference voltage $U_{ref} = 0$, see Fig. (2). The internal resistance of the output of the driving logic gate is represented by the resistance $R$ and the total driven capacitance (given by the sum of the comparator's input capacitance and the parasite capacitances) is represented by the capacitor *C*. The comparator is a logic representation of the input stage of the subsequent gate and we do not deal with the rest of logic operations by that gate. The output of the comparator provides a digital voltage via the signum operation [16,17]: if the input voltage is greater than zero, the output is "*hi*" and if the input voltage is less than zero, the output is "*low*".

When the computer is running, in the case of $U_s(t) = 0$ ("*low*" input), the comparator output shows $p = 0.5$ because the Johnson-noise has zero mean with a symmetric (Gaussian) amplitude density function around zero. However, for $U_s(t) = U_{sH}$ ("*high*" input), the input Johnson noise will be superimposed on the nonzero DC signal $U_{sH}$ thus $p > 0.5$ due to the small asymmetry of the resultant amplitude density function around zero. Note comparator units driven with *analog signals* and additive noise have been used by Stocks [16,17] to propose and demonstrate the so-called *suprathreshold stochastic resonance*, where the noise acts as an information carrier of the analog signal. The noise was band-limited Gaussian white noise.

In this Letter, we avoid any practical question concerning the practical realization of this computer, including the problem of the comparator. We are allowed to do that because we are looking for the lower limit of the energy requirement of processing a single bit. Because the *digital signal must be generated* in the information channel to run the computer, the energy requirement of generating the digital signal at idealistic conditions is an *absolute lower limit* of the energy dissipation.

If we alternate the channel signal between "*low*" and "*high*" levels with the clock frequency, see Fig. (2), then at each clock period we dissipate the charging energy of the capacitor $C$ two times, first when we charge the capacitor and second when we discharge it [7], thus the power dissipation to generate the digital signal is given by:

$$P_s = f_c \frac{1}{2} C U_{sH}^2 \qquad (3)$$

By approximating the Gaussian shape of the amplitude distribution $g(U)$ of the thermal noise with its top value $g(0)$ (in our regime where $p \approx 0.5$) and use the $\sigma = \sqrt{kT/C}$ Johnson relation [7], we get:

$$\Delta p \approx g(0) U_s = \frac{U_s}{\sqrt{2\pi}\,\sigma} = \frac{U_s}{\sqrt{2\pi kT/C}} \ . \qquad (4)$$

Supposing a symmetric time distribution of "low" and "high" bits, the effective $\Delta p$ is the average of the two cases, and from Eqs. (2,3) we obtain:

$$C_{dig}\big|_{p \approx 0.5} = \frac{U_{sH}^2}{\pi \ln 2\ kT/C}\, f_c \ . \qquad (5)$$

Thus, from Eqs. (3,5) we find that the mean energy cost/bit operation is constant:

$$\frac{P_s}{C_{dig}\big|_{p \approx 0.5}} = \frac{\pi \ln 2}{2}\, kT/bit \approx 1.1\ kT/bit \ . \qquad (6)$$

Conversely, the energy efficiency $\eta$ of the data processing is impressive:

$$\eta = \frac{C_{dig}\big|_{p\approx 0.5}}{P_s} = \frac{2}{\pi \ln 2} \; bit/kT \approx 0.9 \; bit/kT \; \approx \; 2.3*10^{20} \; bit/Joule \qquad (7)$$

It is interesting to note that the above results contradict to the general opinion that no digital signal can be used if the energy difference between logic levels is less than $kT*\ln(2)$ or, if not, then a potential barrier of similar height should be between the two states. Such a statement is valid at most for certain types of digital memories. The existence of Shannon's Eq. (1) and the results above indicate that digital signal channels without information storage elements can process information at arbitrary, nonzero energy difference between the logic states.

Today's microprocessors dissipate $>>25,000 \; kT/bit$ energy [8] (though for the low-error operation $\approx 70 \; kT/bit$ would be enough [6-8] so the present $1.1*kT/bit$ value may look promising. However, we should keep in mind that today's Turing type general-purpose computers need error-free operation, and that would require error correcting units (redundancy) and/or error correcting algorithms to function. The error correction need could increase the energy dissipation in the thermal noise driven computer potentially by orders of magnitude and though the information channel capacity would also increase, the resulting $P_s/C_{dig}$ is a non-trivial problem. These results strongly depend on the way of error correction [11-13] and computer architecture.

In conclusion, the main advantage of such a hypothetical thermal noise driven computer would be a potentially improved energy efficiency and an obvious lack of leakage current, cross-talk and ground EMI problems due the very low DC voltages. An apparent disadvantage is the large number of extra (redundancy) elements required for error reduction [11-13].

Finally, we list some of the most important open questions we have not been able to address here:

**1.** *Do we need error correction at all (except input/output operations) when we want to simulate the way the brain works?*

**2.** *Is there any way to realize a non-Turing machine with stochastic elements without excessive hardware/software based redundance?*

**3.** *How should redundancy and error correction be efficiently used to run the thermal noise driven computer as a Turing machine?*

**4.** *How much energy would such error correction cost?*

**5.** *How much energy is needed to feed the comparator?*

**6.** *What is the impact of the noise of the comparator and how to reduce it?*

Though, all these questions are relevant for the ultimate energy dissipation of thermal noise driven computers, the lower limit given in this Letter stays valid because this is the energy need to generate the digital signal.

**Figure caption**

Figure 1.

Shannon's channel capacity of digital channels and the working regimes of the thermal noise driven and classical computers, respectively.

Figure 2.

Model circuitry of the information channel of a possible realization of the thermal noise driven computer. The $U_{th}(t)$ is the inherent Johnson noise voltage of the resistor $R$.

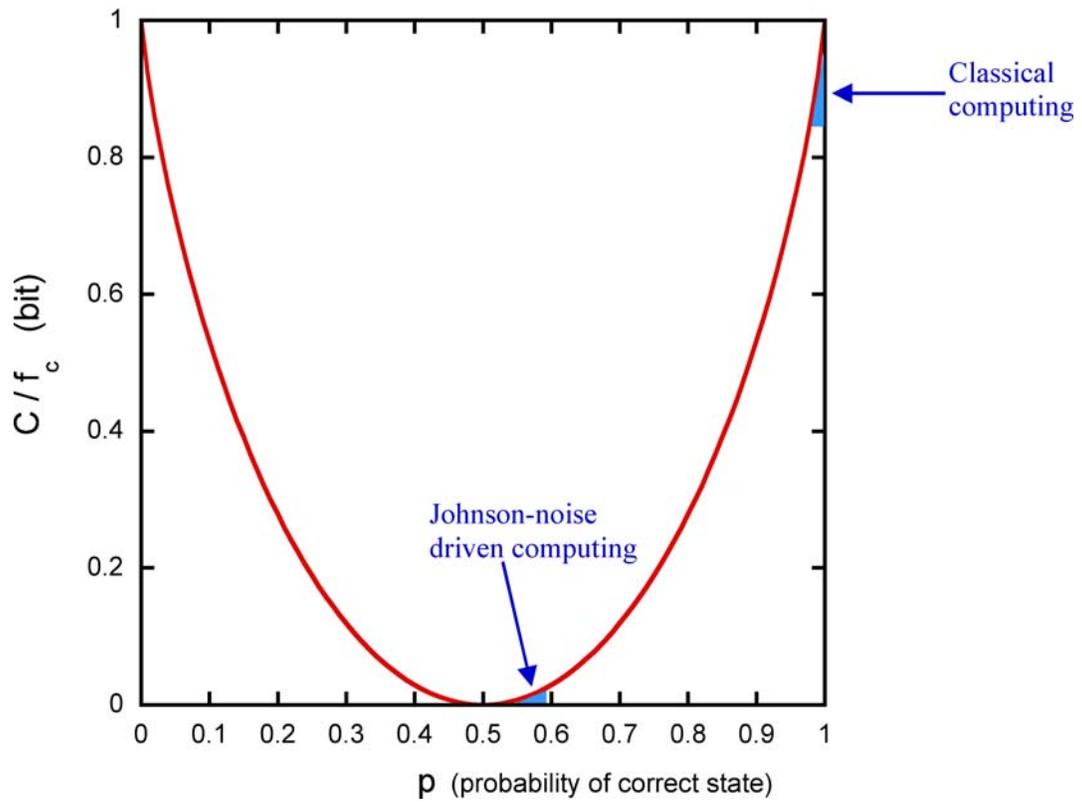

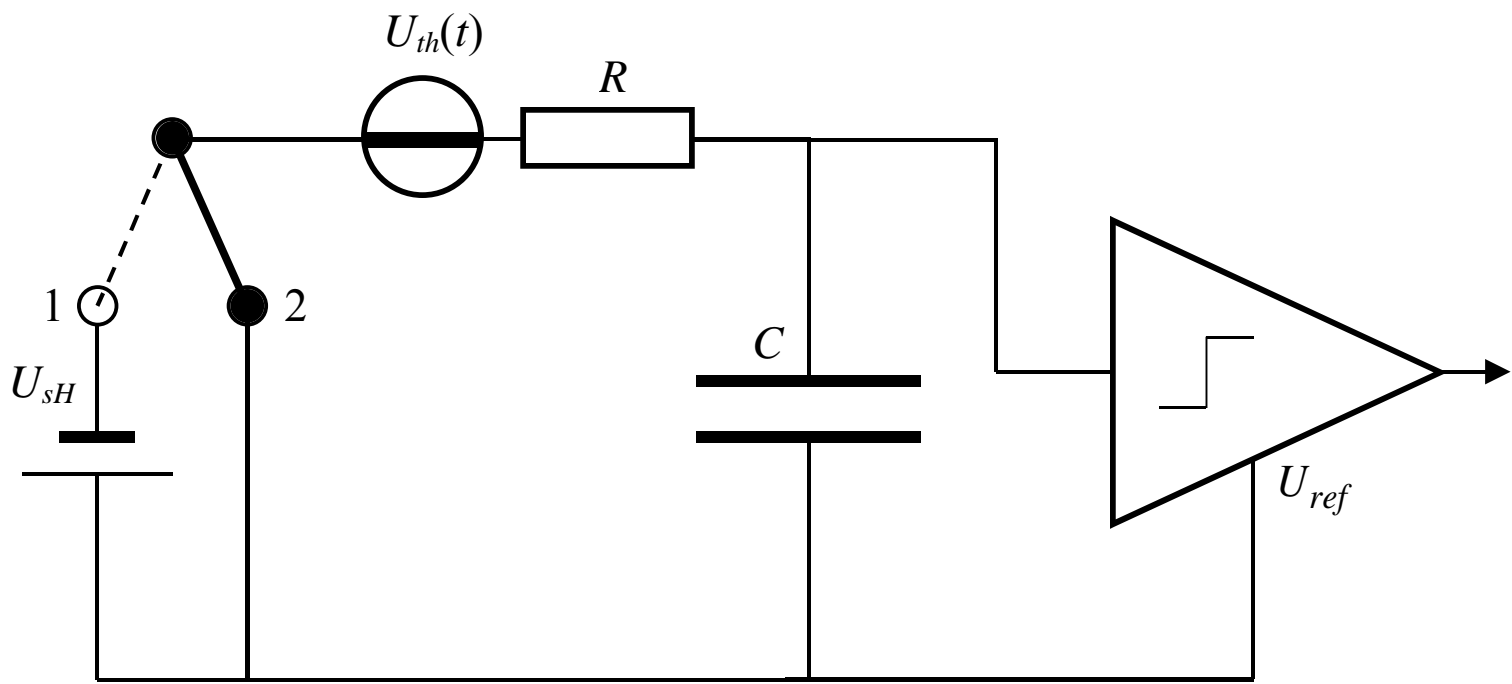